\documentclass[runningheads,leqno]{llncs}
\usepackage[T1]{fontenc}
\usepackage{amsmath}
\usepackage{semantic}
\usepackage{todonotes}
\usepackage{booktabs}
\usepackage{mathtools}
\usepackage{enumitem}
\usepackage{hyperref}
\usepackage{amssymb}
\usepackage{soul}
\usepackage{listings}
\usepackage{color}

\begin{document}
\title{Shape Expressions with Inheritance}


\author{%
Iovka Boneva\inst{1}\orcidID{0000-0002-2696-7303}
\and
Jose Emilio Labra Gayo\inst{2}\orcidID{0000-0001-8907-5348} 
\and
Eric Prud'hommeaux\inst{3}\orcidID{0000-0003-1775-9921}
\and
Katherine Thornton\inst{4}\orcidID{0000-0002-4499-0451}
\and
Andra Waagmeester\inst{5}\orcidID{0000-0001-9773-4008}}
\authorrunning{I. Boneva et al.}
%
\institute{%
Univ. Lille, CNRS, Centrale Lille, UMR 9189 CRIStAL, F-59000 Lille, France
\email{iovka.boneva@univ-lille.fr} (corresponding author)
\and
WESO (WEb Semantics Oviedo) Research Group, University of Oviedo, Spain 
\email{labra@uniovi.es} 
\and 
Janeiro Digital, USA
\email{eric@uu3.org}
\and
Yale University Library, New Haven, CT, USA
\email{katherine.thornton@yale.edu}
\and
Micelio BV, Ekeren, Belgium
\email{andra@micelio.be}
}
\maketitle              
\begin{abstract}
We formally introduce an inheritance mechanism for the Shape Expressions language (ShEx).
It is inspired by inheritance in object-oriented programming languages, and provides similar advantages such as reuse, modularity, and more flexible data modelling.
Using an example, we explain the main features of the inheritance mechanism.
We present its syntax and formal semantics.
The semantics is an extension of the semantics of ShEx 2.1.
It also directly yields a validation algorithm as an extension of the previous ShEx validation algorithms, while maintaining the same algorithmic complexity. 
\keywords{RDF \and schemas for graphs \and formal semantics of languages}
\end{abstract}


\newcommand{\aline}{\textit{line}}
\newcommand{\expr}{\textit{expr}}
\newcommand{\used}{\textit{useful}}
\newcommand{\acc}{\textit{acc}}
\newcommand{\troot}[1]{r_{#1}}
\newcommand{\typing}{\textit{typing}}
\newcommand{\rrepl}[3]{{#1}[#2/#3]}
\newcommand{\PP}{\mathcal{P}}

\newcommand{\algonse}{\textit{sat-n-se}}
\newcommand{\algoMse}{\textit{sat-M-se}}
\newcommand{\algoMte}{\textit{sat-M-te}}

\newcommand{\grayed}[1]{{\color{gray}#1}}
\newcommand{\deprec}[1]{\textcolor{white!70!teal}{#1}}


\newcommand{\extype}[1]{\textit{#1}}
\newcommand{\expred}[1]{\textit{#1}}
\newcommand{\excond}[1]{T_{#1}}

\newcommand{\exFigure}{\extype{Figure}}
\newcommand{\exCircle}{\extype{Circle}}
\newcommand{\exColFig}{\extype{ColouredFigure}}
\newcommand{\exColCircle}{\extype{ColouredCircle}}

\newcommand{\exAttribute}{\extype{Attribute}}
\newcommand{\exCoord}{\extype{Coord}}
\newcommand{\exCoordTwo}{\extype{Coord2D}}
\newcommand{\exCoordThree}{\extype{Coord3D}}
\newcommand{\exColor}{\extype{Colour}}
\newcommand{\exRadius}{\extype{Radius}}
\newcommand{\exProduct}{\extype{Product}}
\newcommand{\exMyProduct}{\extype{MyProduct}}
\newcommand{\exCartesianCoord}{\extype{CartesianCoord}}
\newcommand{\exPolarCoord}{\extype{PolarCoord}}
\newcommand{\exClient}{\extype{Client}}
\newcommand{\exMyClient}{\extype{MyClient}}

\newcommand{\exrgbcolor}{\excond{rgbstring}}
\newcommand{\excolorscope}{\excond{\text{"fill"/"contour"}}}
\newcommand{\exint}{\excond{\mathit{int}}}
\newcommand{\exeven}{\excond{\mathit{even}}}
\newcommand{\exinf}{\excond{<5}}
\newcommand{\exsup}{\excond{>5}}
\newcommand{\exfloat}{\excond{\mathit{float}}}
\newcommand{\exstr}{\excond{\mathit{str}}}
\newcommand{\exany}{\excond{\mathit{any}}}
\newcommand{\extextcontour}{\text{"contour"}}
\newcommand{\extextfill}{\text{"fill"}}
\newcommand{\extextradius}{\text{"radius"}}
\newcommand{\exradius}{\excond{\extextradius}}
\newcommand{\extextcolor}{\text{"colour"}}
\newcommand{\excolor}{\excond{\extextcolor}}
\newcommand{\exredcolor}{\text{"\#ff0000"}}

\newcommand{\exattr}{\expred{attr}}
\newcommand{\exname}{\expred{name}}
\newcommand{\exvalue}{\expred{value}}
\newcommand{\exx}{\expred{x}}
\newcommand{\exy}{\expred{y}}
\newcommand{\exz}{\expred{z}}
\newcommand{\excoord}{\expred{coord}}
\newcommand{\exscope}{\expred{scope}}
\newcommand{\exprecision}{\expred{precision}}
\newcommand{\exradiusprop}{\expred{radius}}
\newcommand{\exreview}{\expred{review}}
\newcommand{\exemail}{\expred{email}}

\newcommand{\iri}{\textsf{IRI}}
\newcommand{\blank}{\textsf{Blank}}
\newcommand{\lit}{\textsf{Literal}}
\newcommand{\triples}{\textsf{Triples}}

\newcommand{\graph}{\mathbf{G}}
\newcommand{\nodes}{\mathit{nodes}}
\newcommand{\subject}{subject}
\newcommand{\neigh}{\mathit{neigh}}


\newcommand{\sch}{\mathbf{S}}
\newcommand{\LabY}{\mathbf{Y}}
\newcommand{\LabX}{\mathbf{X}}
\newcommand{\LabA}{\mathbf{A}}
\newcommand{\LabZ}{\mathbf{Z}}
\newcommand{\sdef}{\mathbf{def}}

\newcommand{\rDef}{{\color{red} \;\Coloneqq\;\; }}
\newcommand{\rSep}{{\color{red} \;\;\big|\;\; }}
\newcommand{\rEnd}{{\color{red} \ .\ }}
\newcommand{\optional}[1]{\textcolor{red}{[} #1  \textcolor{red}{]}}

\newcommand{\teparen}[1]{\langle\, #1 \,\rangle}
\newcommand{\oper}[1]{\mathsf{#1}} 
\newcommand{\andop}{\mathbin{\oper{and}}}
\newcommand{\orop}{\mathbin{\oper{or}}}
\newcommand{\notop}{\mathop{\oper{not}}}
\newcommand{\oneop}{\mathbin{\oper{\mid}}}
\newcommand{\eachop}{\mathbin{\oper{;}}}
\newcommand{\closed}{\mathop{\oper{closed}}}
\newcommand{\extra}{\mathop{\oper{extra}}}
\newcommand{\extends}{\mathop{\oper{extends}}}
\newcommand{\abst}{\mathop{\oper{abstract}}}
\newcommand{\rep}{\oper{*}}
\newcommand{\sref}{\oper{@}}

\newcommand{\nonterm}[1]{\mathit{#1}} 
\newcommand{\se}{\nonterm{s}}
\newcommand{\sse}{\nonterm{u}}
\newcommand{\extse}{\nonterm{t}}
\newcommand{\shape}{\nonterm{h}}
\newcommand{\te}{\nonterm{e}}
\newcommand{\cond}{\nonterm{c}}
\newcommand{\ttc}{\nonterm{f}}
\newcommand{\tc}[2]{#1\; \sref#2}

\newcommand{\anc}{\mathit{anc}}
\newcommand{\desc}{\mathit{desc}}

\newcommand{\AlgoFigures}{Tables~\ref{fig:sem-te},~\ref{fig:sem-se-n} and~\ref{fig:sem-se-M}}
\newcommand{\Correct}{\mathcal{C}}

\newcommand{\Sat}[3]{\graph, #1, #2 \models_\sch #3}
\newcommand{\Satnot}[3]{\graph, #1, #2 \mathbin{\neg\models}_\sch #3}
\newcommand{\sat}[3]{#1, #2 \models #3}
\newcommand{\satnot}[3]{#1, #2 \mathbin{\neg\models} #3}
\newcommand{\satjump}[3]{\sat{#1}{#2}{#3}}

\newcommand{\props}{\mathit{props}}
\newcommand{\tcs}{\mathit{tcs}}
\newcommand{\mainte}{\textit{ext-te}}
\newcommand{\restr}{\mathit{restr}}
\newcommand{\Msat}{M^{\te}}
\newcommand{\Munsat}{M^{\not\te}}
\newcommand{\taumax}{\tau_{\mathrm{max}}}
\newcommand{\stratum}{\sigma}

\newcommand{\AND}{\ \ \ \land\ \ \ }
\newcommand{\OR}{\ \ \ \lor\ \ \ }
\newcommand{\DOT}{.\;\;}
\newcommand{\AFF}{:=}
\newcommand{\NEG}{\neg\;}
\newcommand{\CASE}{\texttt{case}\;}
\newcommand{\WHERE}{\texttt{with}\;}

\newcommand{\subexpr}{\textit{sub-expr}}
\newcommand{\negsubexpr}{\textit{neg-sub-expr}}
\newcommand{\boolsubexpr}{\textit{bool-sub-expr}}

\newcommand{\rulestyle}[1]{{\scriptsize{#1.}}}
\newcommand{\ruleepsilon}{\rulestyle{1}}
\newcommand{\rulefirst}{\ruleepsilon}
\newcommand{\ruletc}{\rulestyle{2}}
\newcommand{\ruleone}{\rulestyle{3}}
\newcommand{\ruleeach}{\rulestyle{4}}
\newcommand{\rulerep}{\rulestyle{5}}

\newcommand{\rulecond}{\rulestyle{6}}
\newcommand{\ruleor}{\rulestyle{7}}
\newcommand{\ruleand}{\rulestyle{8}}
\newcommand{\rulenot}{\rulestyle{9}}
\newcommand{\ruleref}{\rulestyle{10}}
\newcommand{\ruleshape}{\rulestyle{11}}
\newcommand{\ruleextse}{\rulestyle{12}}

\newcommand{\rulecondM}{\rulestyle{13}}
\newcommand{\ruleorM}{\rulestyle{14}}
\newcommand{\ruleandM}{\rulestyle{15}}
\newcommand{\rulenotM}{\rulestyle{16}}
\newcommand{\ruleshapeM}{\rulestyle{17}}
\newcommand{\ruleextseM}{\rulestyle{18}}
\newcommand{\rulelast}{\ruleextseM}
\newcommand{\rulereflM}{\rulestyle{19}}

\newcommand{\depstyle}[1]{\textit{#1}}
\newcommand{\depextends}{\depstyle{dep-extends}}
\newcommand{\depsimple}{\depstyle{dep}}
\newcommand{\depbool}{\depstyle{\deprec{dep-bool}}}
\newcommand{\depshapeneg}{\depstyle{dep-shape-neg}}
\newcommand{\depextraneg}{\depstyle{dep-extra-neg}}

\newcommand{\LabYstratum}{\LabY^{\stratum}}
\newcommand{\LabXstratum}{\LabX^{\stratum}}
\newcommand{\LabAstratum}{\LabA^{\stratum}}
\newcommand{\schstratum}{\sch^{\stratum}}
\newcommand{\sdefstratum}{\sdef^{\stratum}}

\newcommand{\nM}{\frac{M}{n}}

\newcommand{\trestr}[2]{#1_{#2}}
\newcommand{\refbool}{\textbf{\scriptsize ref-bool}}
\newcommand{\refshapepos}{\textbf{\scriptsize ref-shape+}}
\newcommand{\refshapeneg}{\textbf{\scriptsize ref-shape$-$}}
\newcommand{\depext}{\textbf{\scriptsize dep-ext}}
\newcommand{\negref}{\textbf{\scriptsize neg-ref}}
\newcommand{\negextra}{\textbf{\scriptsize neg-extra}}
\newcommand{\dep}{\mathit{dep}}
\newcommand{\labels}{\mathit{labs}}
\newcommand{\ruleoneone}{\ruleone}
\newcommand{\ruleonetwo}{\ruleone}
\newcommand{\ruleorone}{\ruleor}
\newcommand{\ruleortwo}{\ruleor}
\newcommand{\ruleshapeclosed}{\ruleshape}

%



\section{Introduction}

\newcommand{\SHEX}{{shape expression}}
\newcommand{\EX}{{expression}}
\newcommand{\SHPE}{{shape}}

The Shape Expressions language (ShEx) was proposed in 2014 as a concise, high-level language to describe and validate RDF data~\cite{EricSemantics2014}.
It allows us to define ShEx schemas which are collections of \SHPE{}s.
A \SHPE{} describes the required properties of an RDF node and constrains their values.
ShEx 2.1 was published as a W3C final community report in 2019\footnote{\scriptsize\url{https://shex.io/shex-semantics-20191008/}}, and it introduced logical operators on top of shapes, yielding \emph{\SHEX{}s}.
They allow describing alternatives with disjunction, or  further constraining the shape of the data and data values with conjunction.
ShEx has been increasingly adopted in different domains such as HL7 FHIR\footnote{\scriptsize\url{https://www.hl7.org/fhir/}} and Wikidata~\cite{Thornton2019}, where it is used to describe data models and to validate entity data.

The adoption of ShEx for describing large data models has led to the appearance of collaborative shape schema ecosystems like that of Wikidata\footnote{\scriptsize\url{https://www.wikidata.org/wiki/Wikidata:Database_reports/EntitySchema_directory}}.
An important requirement for such ecosystems is to support reusability.
The simplest feature for reuse is to import a \SHEX{} that has been defined elsewhere.
ShEx 2.1 does offer this possibility and the imported \SHEX{} can be reused as is.
Data modelling and reuse often require adapting the existing data and its model to one's particular needs.
Two common adaptation patterns are to \emph{restrict} the allowed values of existing properties, and to \emph{extend} the data model and the data with properties that are specific to a new application.
Additionally, a commonly-used feature is to define frequently occurring patterns which serve as building blocks of \SHEX{}s.

In response to submitted use cases, the ShEx Community Group\footnote{\scriptsize\url{https://www.w3.org/community/shex/}} has proposed an inheritance mechanism for ShEx that provides the afore-mentioned features.
Inspired by inheritance in programming languages, it allows for the definition of a hierarchy of \SHEX{}s.
Child \SHPE{}s can extend their parents with new required properties for RDF nodes, and can impose additional constraints on the properties of their parents.
Additionally, as in object-oriented programming languages, a node with a child \SHPE{} can be used where a parent \SHPE{} is expected. 
Multiple inheritance is also allowed.
Finally, \SHEX{}s can be made \emph{abstract} indicating that they cannot be used on their own, but only as building blocks of other \EX{}s.
An early version of the inheritance mechanism has been used in~\cite{Sharma2023} to translate in ShEx the user-facing documentation of data structures in FHIR.

In this paper, we introduce the formal semantics for ShEx with inheritance, built as an extension of the formal semantics of ShEx~2.1~\cite{Boneva17}.
The formalization posed some challenges, notably for handling multiple inheritance.
In particular, we introduce a syntactic restriction on the combined use of inheritance with disjunction.
The restriction avoids unnecessary complexity in the language, while still providing the functionality required by the use cases.
The ShEx formalization presented here will be submitted for the next version of the ShEx language, currently under standardization by IEEE.\footnote{\scriptsize\url{https://standards.ieee.org/ieee/3330/11119/}}

The paper is organized as follows. 
In Sect.~\ref{sec:motivating-example} we give an example illustrating the main features of the inheritance mechanism, and define some preliminary notions.
Sect.~\ref{sec:syntax} gives the syntax, and Sect.~\ref{sec:semantics} the semantics of the language.
Sect.~\ref{sec:algorithm} presents a validation algorithm.
We close with a discussion, an overview of related work and a conclusion in Sect.~\ref{sec:discussion}.
The proofs omitted in the paper are presented in appendix.
A companion webpage~\cite{companion-github-repo} lists the current implementations and provides source code and demonstrations for the examples.


\section{Motivating example and background}
\label{sec:motivating-example}

\begin{figure}[b]
	\centering
	\fontsize{8}{9}
\begin{tikzpicture}[>=latex,scale=1.1]
\node[outer sep=2pt] (r1) at (0,0) {\tt f1};
\node[outer sep=2pt] (r1c) at (1.2,-0.5) {\tt c1};
\node (r1cx) at (2.6,-0.3) {2.0};
\node (r1cy) at (2.6,-0.7) {4.0};

\node (r1w) at (1,0.8) {\tt a1};
\node (r1wn) at (2.6,0.9) {\extextradius};
\node (r1wv) at (2.6,0.4) {10.1};


\node (r1col) at (-1.2,0) {\tt a2};
\node (r1colv) at (-2.4,0.8) {\exredcolor};
\node (r1coln) at (-2.6,-0.7) {\extextcolor};
\node (r1cols) at (-3,0) {\extextfill};

\draw (r1) edge[->] node[midway, sloped, below] {\excoord} (r1c);
\draw (r1c) edge[->] node[midway,above] {\exx} (r1cx);
\draw (r1c) edge[->] node[midway,below] {\exy} (r1cy);
\draw (r1) edge[->] node[midway,above,sloped] {\exattr} (r1w);
\draw (r1) edge[->] node[midway,above] {\exattr} (r1col);
\draw (r1w) edge[->] node[midway, above,sloped] {\exname} (r1wn);
\draw (r1w) edge[->] node[midway, below,sloped] {\exvalue} (r1wv);
\draw (r1col) edge[->] node[pos=0.4,below,sloped] {\exname} (r1coln);
\draw (r1col) edge[->] node[above,sloped] {\exvalue} (r1colv);
\draw (r1col) edge[->] node[midway,above] {\exscope} (r1cols);

\node[outer sep=2pt] (r2) at (4,0) {\tt f2};
\node[outer sep=2.pt] (r2c) at (5.1,-0.5) {\tt c2};
\node (r2cx) at (6.5,-0.3) {0.2};
\node (r2cy) at (6.5,-0.7) {-2.3};

\node[outer sep=2pt] (r2w) at (5.1,0.8) {\tt a3};
\node (r2wn) at (6.7,0.9) {\extextradius};
\node (r2wv) at (6.5,0.4) {7.2};


\draw (r2) edge[->] node[midway, sloped, below] {\excoord} (r2c);
\draw (r2c) edge[->] node[midway,above] {\exx} (r2cx);
\draw (r2c) edge[->] node[midway,below] {\exy} (r2cy);

\draw (r2) edge[->] node[midway,above, sloped] {\exattr} (r2w);

\draw (r2w) edge[->] node[midway, above,sloped] {\exname} (r2wn);
\draw (r2w) edge[->] node[midway, below,sloped] {\exvalue} (r2wv);

\end{tikzpicture}
	\caption{Example graph with nodes {\tt f1}, {\tt f2}, {\tt a1}, {\tt a2}, {\tt a3} and strings and float literals.}
	\label{fig:example-graph}
\end{figure}
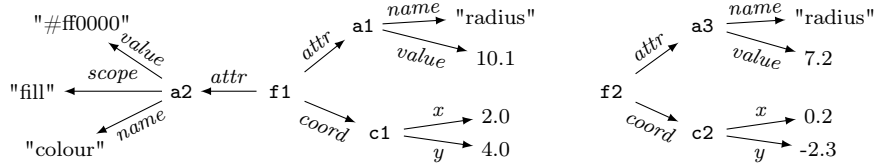

Assume the following constraints on nodes in an RDF graph: $\exstr$ is satisfied by string literals, $\exfloat$ by float literals, $\exany$ by all RDF nodes,
$\excolor$ is satisfied by the literal \extextcolor, and $\exradius$ by the literal \extextradius.
A ShEx schema using inheritance is presented in Fig.~\ref{fig:example-attribute-color}, we will describe it section by section. 
The source code for that schema is available on the companion webpage \cite{companion-github-repo}.
The \SHEX{} named \exCoord{} describes nodes representing coordinates, i.e. having properties $x$ and $y$ whose values are floats.
The \SHEX{} \exAttribute{} requires a property \exname{} which is a string, and a property \exvalue{} that can be anything.
The $\extends \_$ indicates that the \SHEX{} is being extended. 
A \exColor{} is a specific kind of attribute (extends on \exAttribute) that additionally to the \exname{} and \exvalue{} properties requires a \exscope{} property whose value is a string.
This is captured by the first part of the definition, preceding the $\andop$ keyword.
The second part states that the \exname{} of a \exColor{} attribute is \extextcolor.
Thus, inheritance provides a mechanism for requiring additional properties (before the $\andop$) and restricting existing properties (after the $\andop$). 
Fig.~\ref{fig:example-graph} shows an example graph.
Nodes {\tt a1}, {\tt a2}, {\tt a3} satisfy \exAttribute{}, node {\tt a2} satisfies \exColor, and node {\tt c1} satisfies \exCoord.

\begin{figure}[t]
$$
\begin{array}{rl}
\exCoord &\to \teparen{ \tc{\exx}{\exfloat} \;\eachop\; \tc{\exy}{\exfloat} }\\
\exAttribute &\to \extends \_ \teparen{\tc{\exname}{\exstr} \eachop \tc{\exvalue}{\exany}} \\
\exColor &\to \extends\;\exAttribute\; \teparen{\tc{\exscope}{\exstr}} \;\;\andop\;\; \teparen{\tc{\exname}{\excolor}}\\[\medskipamount]
\abst \exFigure &\to \extends \_ \teparen{\tc{\excoord}{\exCoord}} \\
\exCircle &\to \extends\; \exFigure\; \teparen{\tc{\exattr}{\exRadius}}\\
\exRadius &\to \extends\;\exAttribute\; \teparen{\varepsilon} \andop \teparen{\tc{\exname}{\exradius} \;\eachop\; \tc{\exvalue}{\exfloat}}\\[\medskipamount]
\exColFig &\to \extends\;\exFigure\; \teparen{\tc{\exattr}{\exColor}}\\
\exColCircle &\to \extends\;\exCircle, \exColFig\; \teparen{\varepsilon}
\end{array}
$$
\caption{A ShEx schema with inheritance.}
\label{fig:example-attribute-color}
\label{fig:example-figures-coordinates}
\label{fig:example-multiple-inheritance}
\end{figure}

A \exFigure{} has coordinates associated with the property \excoord.
The \SHEX{} \exFigure{} is marked as $\abst$, which indicates that it cannot be directly satisfied, rather one of its non-abstract descendants will be satisfied.
A \exCircle{} is a \exFigure{} that has a \exRadius{} attribute reachable through the property \exattr, which in turn is an \exAttribute{} having \exname{} \extextradius{} and a float \exvalue{}.
Note that the definition of \exCircle{} does not have a restriction after the $\andop$; in fact the restriction is optional.
On the other hand, the extending portion of \exRadius{} is $\varepsilon$, meaning that a \exRadius{} does not require additional properties beyond those specified in its parent \exAttribute.
In the example graph, {\tt f2} is a \exCircle{}, {\tt a1}, {\tt a3} satisfy \exRadius{}.

The final part of the example illustrates multiple inheritance.
A \exColCircle{} extends both on \exColFig{} and \exCircle{} which in turn both extend from \exFigure.
Recall that a \exFigure{} has one \excoord{} property, which is inherited twice by \exColCircle.
The inheritance mechanism ensures that multiply-inherited properties are not used more than once.
In this particular example, a \exColCircle{} can have only one \excoord. 
In the example graph, {\tt f1} is a \exColCircle{}.

\begin{figure}[tb]
\centering
\begin{tikzpicture}[>=latex]
\node (attr) at (-4,0.7) {\exAttribute};
\node (color) at (-5,0) {\exColor};
\node (width) at (-3.2,0) {\exRadius};

\draw (color) edge[->] (attr);
\draw (width) edge[->] (attr);

\node (fig) at (-0.2,0.7) {\exFigure};
\node (rect) at (-1.4, 0) {\exCircle};
\node (colfig) at (0.9,0) {\exColFig};
\node (colrect) at (-0.2, -0.7) {\exColCircle};
	
\draw (rect) edge[->] (fig);
\draw (colfig) edge[->] (fig);
\draw (colrect) edge[->] (rect);	
\draw (colrect) edge[->] (colfig);


\end{tikzpicture}
\caption{Extension hierarchy graph for the example.}
\label{fig:example-inheritance-graph}
\end{figure}
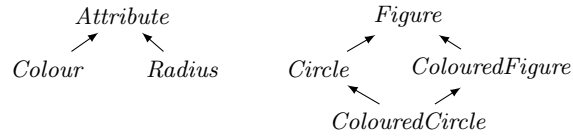

As in object-oriented programming languages, inheritance yields an extension hierarchy which is a directed acyclic graph. 
Fig.~\ref{fig:example-inheritance-graph} depicts the inheritance graph of the example schema, with arrows going from \SHEX{}s to their direct ancestors. 
The inheritance mechanism requires that if a node satisfies some of the descendants of a \SHEX{}, then it satisfies the \SHEX{} itself.
Therefore, {\tt f1} is a \exColCircle, a \exColFig{}, a \exCircle{}, and a \exFigure.

\subsubsection{Background}

\iri, \blank{} and \lit{} are three countable mutually disjoint sets of IRIs, blank nodes, and literals, respectively.
Elements of $(\iri \cup \blank) \times \iri \times (\iri \cup \lit \cup \blank)$ are called triples.
If $(n, p, o)$ is a triple, then $n$ is its subject, $p$ its property, and $o$ its object.
An RDF graph $\graph$ is a finite set of triples.
The set of nodes of $\graph$ is the set of all subjects and objects of its triples and is denoted $\nodes(\graph)$.
%
Given a graph $\graph$ and a node $n \in \nodes(\graph)$, the \emph{neighbourhood} of $n$ in $\graph$ is the set $\neigh_\graph(n)$ of triples having $n$ as subject. 
Formally, $\neigh_\graph(n) = \{(n, p, o) \mid (n, p, o) \in \graph\}$.
A set of triples $M \subseteq \triples$ is called a \emph{neighbourhood set of triples} if all its elements have the same subject denoted $\subject(M)$.

We use the symbol $\uplus$ for disjoint union, i.e. $M = M_1 \uplus \cdots \uplus M_k$ (for some $k \in \bbbn$) means that $M = M_1 \cup \cdots \cup M_k$ and $M_i \cap M_j = \emptyset$ whenever $i \neq j$.
The disjoint union of zero sets is the empty set, i.e., if $k = 0$, then $M = \emptyset$. In grammars or expressions, square brackets $\optional{\;}$ surround optional elements.


\section{Syntax}
\label{sec:syntax}

A ShEx schema consists of a set of labeled (named) \SHEX{}s to be checked for conformance with the nodes of an RDF graph.
We start by defining \SHEX{}s, then give a formal definition of ShEx schemas with inheritance and the corresponding extension hierarchy.

\begin{definition}[shape expressions]
	\label{def:shape-expr}
	Let $\LabY$ be a finite set of \emph{simple labels} and $\LabX$ be a finite set of \emph{extendable labels} disjoint from $\LabY$. 
	A \emph{shape expression} over $\LabY$ and $\LabX$ is either a \emph{reference} of the form $\sref z$, or a \emph{plain shape expression} derivable from the non-terminal $\sse$ in the following grammar:
	\begin{align*}
	\sse & \rDef \shape \rSep \extse \rSep \cond \rSep \sse\, \orop \sse \rSep \sse\, \andop \sse \rSep \notop\, \sse \rEnd && \text{(plain shape expression)} \\
	\shape & \rDef \optional{\closed}\;\; \optional{\extra\, P}\;\; \teparen{\te} \rEnd && \text{(shape)}\\
	\extse & \rDef \extends X\ \shape \rEnd && \text{(shape with extends)} \\
	\cond & \rDef \text{a boolean function over } \iri \cup \blank \cup \lit \rEnd && \text{(node constraint)}\\
	\te & \rDef \varepsilon \rSep \tc{p}{z} \rSep \te \oneop \te \rSep \te \eachop \te \rSep \te\rep \rEnd&& \text{(triple expression)}
\end{align*}
	where $z \in \LabY \cup \LabX$, $X \subseteq \LabX$, $p \in \iri$, $P \subseteq \iri$, the symbol `$\varepsilon$' is a constant, and the symbols `$\sref$', `$\oneop$', `$\rep$' and `$\eachop$' are operators.
	An expression derivable by the non-terminal $\shape$, resp. $\extse$, resp. $\cond$, resp. $\te$, is called a \emph{shape}, resp. a \emph{shape with extends}, resp. a \emph{node constraint}, resp. a \emph{triple expression}.
	An expression of the form $\tc{p}{z}$ is called a \emph{triple constraint}.
\end{definition}

A shape with extends is a new kind of \SHEX{}. It is constructed from a (possibly empty) set $X$ of extendable labels and a shape $\shape$.
The labels in $X$ indicate which \SHEX{}s are being extended.
All other ingredients of \SHEX{}s are present in ShEx~2.1.

\begin{definition}[ShEx schema]
\label{def:schema}
A \emph{ShEx schema (with inheritance)} is a tuple $\sch = (\LabY, \LabX, \sdef, \LabA)$, where $\LabY$ is a finite set of \emph{simple labels}, $\LabX$ is a finite set of \emph{extendable labels} disjoint from $\LabY$, $\LabA \subseteq \LabX$ are the \emph{abstract} labels, and $\sdef$ is a function in which every label from $\LabY \cup \LabX$ associates its definition, with:
\begin{itemize}
\item if $y \in \LabY$, then $\sdef(y)$ is a plain shape expression;
\item if $x \in \LabX$, then $\sdef(x)$ is a shape expression of the form $\extends\, X\; h\, \optional{\andop\, \sse}$, also called an \emph{extendable shape expression}, where $\shape$ is a shape and $\sse$ is a plain shape expression.
\end{itemize}
\end{definition}

\begin{example}
	\label{ex:explanation-example-syntax}
	In Sect.~\ref{sec:motivating-example} we presented a schema -- denote it $\sch_1 = (\LabY_1, \LabX_1, \sdef_1, \LabA_1)$ -- as a set of rules of the form $z \rightarrow s$, to be understood as $\sdef_1(z) = s$.
	The simple labels are $\LabY_1 = \left\{\exCoord, \exany, \exstr, \exfloat,  \excolor, \exradius\right\}$.
	The definitions of the sub-scripted $\excond{}$'s correspond to the node constraints described in the text, e.g., $\sdef_1(\exstr)$ is the function that is true for string literals and false for all other nodes. 
	The extendable labels are those that appear on the left-hand sides of the rules, except for \exCoord{}, i.e.$\LabX_1 = \left\{\exAttribute, \exColor, \ldots, \exColCircle \right\}$.
	Their definitions are extendable shape expressions, in which we omit the curly braces around the elements of the set $X$ and write $\_$ if it is empty.
	Finally, $\LabA_1 = \{\exFigure\,\}$ is indicated by the \texttt{abstract} keyword.
\end{example}
In the sequel, all examples of schemas will use the same notation as the one used in Sect.~\ref{sec:motivating-example} and explained in Example~\ref{ex:explanation-example-syntax}.

\begin{definition}[extension hierarchy]
\label{def:ext-hier}
	The \emph{extension hierarchy} of the schema $\sch = (\LabY, \LabX, \sdef,  \LabA)$ is a directed graph denoted $H_\sch$ whose set of nodes is $\LabX$ and that has an edge from $x$ to $x'$ if and only if $\sdef(x) = \extends X\; h\; \optional{\andop \sse}$ and $x' \in X$. 
	If there exists a (possibly empty) path from $x$ to $x'$ in $H_\sch$, we say that $x'$ is an \emph{ancestor} of $x$ or, equivalently, that $x$ is a \emph{descendant} of $x'$.
	We denote $\anc(x)$, resp. $\desc(x)$, the set of ancestors, resp. descendants, of the label $x$.
\end{definition}

\begin{example}
	\label{ex:extension-hierarchy}
	Fig.~\ref{fig:example-inheritance-graph} shows $H_{\sch_1}$, the extension hierarchy of the schema $\sch_1$ from Example~\ref{ex:explanation-example-syntax}.
	Note that the extension hierarchy is determined only by the first shape with extends in the definition of an extendable shape label.
	For instance, let $\sdef(x) = \extends\, x_1, x_2\, \teparen{\varepsilon} \ \andop\  \extends\, x_3, x_4\, \teparen{\varepsilon}$.
	According to Def.~\ref{def:ext-hier}, $x_1$ and $x_2$ are ancestors of $x$, but $x_3$ and $x_4$ are not ancestors of~$x$.
\end{example}

\section{Semantics}
\label{sec:semantics}

Let $\graph$ be a graph and $\sch = (\LabY, \LabX, \sdef, \LabA)$ be a ShEx schema.
A \emph{typing of $\graph$ by $\sch$} is a subset of $\nodes(\graph) \times (\LabY \cup \LabX)$.
We use typings to define the semantics of triple expressions and shape expressions. 
Intuitively, a typing can be understood as a set of assertions about which nodes of the graph satisfy which shape expressions. 
It can be used to derive other such assertions.
In Sect.~\ref{sec:satisfaction-rules} we define the semantics of expressions under a given typing.
A typing $\tau$ is \emph{correct} (w.r.t. a graph and a schema) if each of the assertions it contains can be derived from $\tau$ itself.
In Sect.~\ref{sec:semantics-well-defined} we introduce well-defined schemas by forbidding some circular dependencies between shape expressions.
Finally, in Sect.~\ref{sec:maximal-typing} we show that if a schema is well-defined, then there exists a unique maximal correct typing $\taumax$ which defines the semantics of ShEx schemas, in the sense that it contains all correct assertions about which nodes satisfy which shape expressions.

\subsection{Satisfying an expression under given typing}
\label{sec:satisfaction-rules}

\newcommand{\nOneTwo}{(n, p, 2)}
\newcommand{\nOneFour}{(n, p, 4)}
\newcommand{\nTwoTwo}{(n, p, 2)}
\newcommand{\nTwoA}{(n, p, \text{"a"})}
\newcommand{\nThreeTwo}{(n, p, 2)}
\newcommand{\nThreeFour}{(n, p, 4)}
\newcommand{\nThreeSix}{(n, p, 6)}
\newcommand{\nFourTwo}{(n, p, 2)}
\newcommand{\nFourFour}{(n, p, 4)}
\newcommand{\nFourA}{(n, p, \text{"a"})}
\newcommand{\nQtwo}{(n, q, 2)}

\newcommand{\tauTwoEven}{(2, \exeven)}
\newcommand{\tauFourEven}{(4, \exeven)}
\newcommand{\tauSixEven}{(6, \exeven)}
\newcommand{\tauTwoInf}{(2, \exinf)}
\newcommand{\tauFourInf}{(4, \exinf)}
\newcommand{\tauSixSup}{(6, \exsup)}
\newcommand{\tauAStr}{(\text{"a"}, \exstr)}

\newcommand{\twofour}{{24}}
\newcommand{\twoa}{{2\text{a}}}
\newcommand{\twofoursix}{{246}}
\newcommand{\twofoura}{{24\text{a}}}

\newcommand{\TABLETRIPLEEXPR}[1]{%
\begin{table}[#1]
	\caption{Definition of the satisfiability relation $\sat{M}{\tau}{\te}$}
	\label{fig:sem-te}
	\centering
	\begin{tabular}{rll}
	\toprule
	& $\te$ & $\sat{M}{\tau}{\te}$\\
	\midrule
	\ruleepsilon & $\varepsilon$ & $M = \emptyset$\\
	\ruletc & $\tc{p}{z}$ & $M = \{(n, p, o)\} \AND \satjump{o}{\tau}{\sref z}$\\
	\ruleone & $\te_1 \oneop \te_2\quad$ & $\sat{M}{\tau}{\te_1} \OR \sat{M}{\tau}{\te_2}$ \\ 
	\ruleeach & $\te_1 \eachop \te_2$ & $\exists M_1, M_2\DOT M = M_1 \uplus M_2 \AND  \sat{M_1}{\tau}{\te_1} \AND \sat{M_2}{\tau}{\te_2}$\\
	\rulerep & $\te_1\rep$  & $\exists k \in 0..|M|\DOT \exists M_1,\ldots,M_k\DOT$ \\
	&& \qquad $M = M_1 \uplus \cdots \uplus M_k \AND \forall i \in 1..k\DOT \sat{M_i}{\tau}{\te_1}$\\
	\bottomrule
\end{tabular}

\end{table}}

\newcommand{\TABLESHAPEXPRS}[1]{%
\begin{table}[#1]
	\caption{Definition of the satisfiability relation $\sat{n}{\tau}{\se}$.}
	\label{fig:sem-se-n}
	\centering
\begin{tabular}{rll}
\toprule
& $\se$ & $\sat{n}{\tau}{\se}$\\
\midrule
\rulecond & $\cond$ & $\cond(n)$ \\
\ruleor & $\se_1 \orop \se_2$ &  $\sat{n}{\tau}{\se_1} \OR \sat{n}{\tau}{\se_2}$\\
\ruleand & $\se_1 \andop \se_2$ & $\sat{n}{\tau}{\se_1} \AND \sat{n}{\tau}{\se_2}$\\
\rulenot & $\notop \se_1$ & $\satnot{n}{\tau}{\se_1}$\\
\ruleref & $\sref z$ & $(n, z) \in \tau$ \\
\ruleshape & $\optional{\closed}\ \optional{\extra P}\ \teparen{\te}$\hspace{0.5cm} & $\satjump{\neigh_\graph(n)}{\tau}{\optional{\closed}\ \optional{\extra P}\ \teparen{\te}}$  \\
\ruleextse & $\extends X\; \shape$ & $\satjump{\neigh_\graph(n)}{\tau}{\extends X\; \shape}$ \\
\bottomrule
\end{tabular}

	\end{table}}

\newcommand{\TABLESHAPEEXPRM}[1]{%
\begin{table}[#1]
	\caption{Definition of the satisfiability relation $\sat{M}{\tau}{\se}$.}
	\label{fig:sem-se-M}
	\centering
\begin{tabular}{rlll}
	\toprule
	&$\se$ & $\sat{M}{\tau}{\se}$\\
	\midrule
	\rulecondM &$\cond$ & $\cond(\subject(M))$ \\
	\ruleorM &$\se_1 \orop \se_2$ & $\sat{M}{\tau}{\se_1} \OR \sat{M}{\tau}{\se_2}$\\
	\ruleandM &$\se_1 \andop \se_2$ & $\sat{M}{\tau}{\se_1} \AND \sat{M}{\tau}{\se_2}$\\
	\rulenotM &$\notop \se_1$ & $\satnot{M}{\tau}{\se_1}$\\
	\specialrule{0pt}{0.5pt}{\belowrulesep}
	\ruleshapeM &$\optional{\closed}\optional{\extra P}\teparen{\te}$ 
	& $\sat{\Msat}{\tau}{\te} \AND \props(\Munsat) \subseteq P' \AND $ \\
	&& $\closed \text{ is present} \implies \props(M) \subseteq \props(\te) \cup P'
	$\\
	&& \text{with}\\
	&&$\quad P' \AFF P\ \text{ if } \extra \text{ is present},\  P' \AFF \emptyset\ \text{ otherwise }$\\
	&& $\quad \Msat \AFF \left\{(n, p, o) \in M \mid \exists \tc{p}{z} \in \tcs(\te).\; \satjump{o}{\tau}{\sref z}\right\}$\\
	&&$\quad \Munsat \AFF \left\{(n, p, o) \in M \mid p \in \props(\te)\right\} \setminus \Msat$ \\
	\specialrule{0pt}{0.7pt}{\belowrulesep}
\ruleextseM & $\extends X\; \shape$ &
$\exists M'\; \text{ and }\;\ \exists M_x \text{ for every $x \in \anc(X)$} \text{ such that}$\\
&&$\quad M = M' \uplus \biguplus_{x \in \anc(X)} M_x \AND$\\
&&$\quad \satjump{M'}{\tau}{\shape} \AND$ \\
&&$\quad \text{for every } x \in \anc(X)\DOT$\\ &&$\quad\quad \satjump{M_x}{\tau}{\mainte(x)} \AND$ \\
&&$\quad\quad \restr(x) \text{ present} \implies \satjump{\left(\bigcup_{z \in \anc(x)} M_{z}\right)\!\!}{\tau}{\restr(x)}$ \\
	\bottomrule
\end{tabular}

\end{table}}

\newcommand{\TECHNICALEXAMPLEDATA}{%
\begin{align*}
	& M_\twofour = \{\nOneTwo, \nOneFour\} \qquad \qquad M_\twofoursix = \{\nThreeTwo, \nThreeFour, \nThreeSix\} 
	\\
	& M_\twoa = \{\nTwoTwo, \nTwoA\} \qquad \quad M_\twofoura = \{\nFourTwo, \nFourFour, \nFourA\} \\
	& \tau = \left\{ \tauTwoEven, \tauTwoInf, \tauFourEven, \tauFourInf, \tauSixEven, \tauSixSup, \tauAStr \right\}
\end{align*}}

The semantics of shape expressions and triple expressions are given by the means of three mutually recursive satisfiability relations.
For triple expressions, the satisfiability relation is of the form 
$\Sat{M}{\tau}{\te}$ 
and for shape expressions we use two satisfiability relations, one of the form $\Sat{n}{\tau}{\se}$ and another one of the form $\Sat{M}{\tau}{\se}$ (where
$\sch$ is a schema, $\graph$ is a graph, $n$ is a node of $\graph$, $\tau$ is a typing over $\graph$ and $\sch$, $M$ is a set of neighbourhood triples from $\graph$,
$\te$ is a triple expression, and $\se$ is a shape expression).
The different signatures of the three relations will always allow distinguishing them.
We will omit $\graph$ and $\sch$ whenever they are clear from the context. 
The satisfiability relation of the form $\Sat{M}{\tau}{\se}$ is new for ShEx with inheritance. 
It is necessary for the definition of the semantics of shapes with extends.

For the remainder of the section, consider given a graph $\graph$ and a ShEx schema $\sch = (\LabY, \LabX, \sdef, \LabA)$ that will be implicit (omitted) in all satisfiability relation statements.
The semantics of the satisfiability relations are given in~\AlgoFigures{} inductively on the structure of expressions.
The right-hand side column in each table is a boolean expression giving the truth value for the case from the left-hand side column.
The symbols $\neg$, $\land$, $\lor$, $\exists$ and $\implies$ have their usual meaning from logic. 
The symbol $=$ is test of equality, while $\AFF$ is assignment.
When $\Sat{M}{\tau}{\te}$, resp. $\Sat{n}{\tau}{\se}$, resp. $\Sat{M}{\tau}{\se}$ holds, we say that $M$ \emph{satisfies} $\te$, resp. $n$ \emph{satisfies} $\se$, resp. $M$ \emph{satisfies} $\se$ under typing $\tau$.
In the sequel of this section we explain the satisfiability relations. 

\paragraph{Triple expressions (Table~\ref{fig:sem-te})}
\begin{table}[t]
	\caption{Definition of the satisfiability relation $\sat{M}{\tau}{\te}$}
	\label{fig:sem-te}
	\centering
	\begin{tabular}{rll}
	\toprule
	& $\te$ & $\sat{M}{\tau}{\te}$\\
	\midrule
	\ruleepsilon & $\varepsilon$ & $M = \emptyset$\\
	\ruletc & $\tc{p}{z}$ & $M = \{(n, p, o)\} \AND \satjump{o}{\tau}{\sref z}$\\
	\ruleone & $\te_1 \oneop \te_2\quad$ & $\sat{M}{\tau}{\te_1} \OR \sat{M}{\tau}{\te_2}$ \\ 
	\ruleeach & $\te_1 \eachop \te_2$ & $\exists M_1, M_2\DOT M = M_1 \uplus M_2 \AND  \sat{M_1}{\tau}{\te_1} \AND \sat{M_2}{\tau}{\te_2}$\\
	\rulerep & $\te_1\rep$  & $\exists k \in 0..|M|\DOT \exists M_1,\ldots,M_k\DOT$ \\
	&& \qquad $M = M_1 \uplus \cdots \uplus M_k \AND \forall i \in 1..k\DOT \sat{M_i}{\tau}{\te_1}$\\
	\bottomrule
\end{tabular}

\end{table}
The semantics of triple expressions are the same as those of ShEx~2.1.
Only the empty set of triples satisfies the empty expression $\varepsilon$.
A triple constraint $\tc{p}{z}$ is satisfied by a singleton set which unique triple has $p$ as property and an object that satisfies $\sref z$.
The expression $\te_1 \oneop \te_2$ is satisfied if one of $\te_1$ or $\te_2$ is satisfied.
As for $\te_1 \eachop \te_2$, it is satisfied by $M$ if $M$ can be split in two parts $M_1$ and $M_2$ that satisfy the sub-expressions $\te_1$ and $\te_2$, respectively.
Finally, the repeated expression $\te\rep$ is satisfied by $M$ if $M$ can be split into $k$ parts, for some $k$ less than or equal to the number of elements in $M$, and each of them satisfies $\te$.
Note that the empty set always satisfies $\te\rep$.


\begin{example}
	\label{ex:sat-triple-expr}
	Let $\tau$ be the typing and $M_\twofour, M_\twofoursix, M_\twoa, M_\twofoura$ be the sets of triples defined below. 
	Consider the triple expression $\te = \tc{p}{\exeven} \eachop \left(\tc{p}{\exinf} {}^\rep \oneop \tc{p}{\exstr}\right)$.
	Then $\sat{M_\twofour}{\tau}{\te}$, $\sat{M_\twofoursix}{\tau}{\te}$, $\sat{M_\twoa}{\tau}{\te}$, but $\satnot{M_\twofoura}{\tau}{\te}$.
	\TECHNICALEXAMPLEDATA
\end{example}

\paragraph{Shape expression satisfied by a node (Table~\ref{fig:sem-se-n})}
\begin{table}[t]
	\caption{Definition of the satisfiability relation $\sat{n}{\tau}{\se}$.}
	\label{fig:sem-se-n}
	\centering
\begin{tabular}{rll}
\toprule
& $\se$ & $\sat{n}{\tau}{\se}$\\
\midrule
\rulecond & $\cond$ & $\cond(n)$ \\
\ruleor & $\se_1 \orop \se_2$ &  $\sat{n}{\tau}{\se_1} \OR \sat{n}{\tau}{\se_2}$\\
\ruleand & $\se_1 \andop \se_2$ & $\sat{n}{\tau}{\se_1} \AND \sat{n}{\tau}{\se_2}$\\
\rulenot & $\notop \se_1$ & $\satnot{n}{\tau}{\se_1}$\\
\ruleref & $\sref z$ & $(n, z) \in \tau$ \\
\ruleshape & $\optional{\closed}\ \optional{\extra P}\ \teparen{\te}$\hspace{0.5cm} & $\satjump{\neigh_\graph(n)}{\tau}{\optional{\closed}\ \optional{\extra P}\ \teparen{\te}}$  \\
\ruleextse & $\extends X\; \shape$ & $\satjump{\neigh_\graph(n)}{\tau}{\extends X\; \shape}$ \\
\bottomrule
\end{tabular}

	\end{table}
A node constraint $\cond$ is satisfied by $n$ if $\cond(n)$ holds.
The semantics of the boolean operators $\orop$, $\andop$ and $\notop$ are as expected.
A reference $\sref z$ is satisfied by node $n$ if the typing contains the pair $(n, z)$.
Both a shape and a shape with extends are satisfied by a node if the node's neighbourhood satisfies the shape.

\paragraph{Shape expression satisfied by a set of triples (Table~\ref{fig:sem-se-M}).}
\begin{table}[t]
	\caption{Definition of the satisfiability relation $\sat{M}{\tau}{\se}$.}
	\label{fig:sem-se-M}
	\centering
\begin{tabular}{rlll}
	\toprule
	&$\se$ & $\sat{M}{\tau}{\se}$\\
	\midrule
	\rulecondM &$\cond$ & $\cond(\subject(M))$ \\
	\ruleorM &$\se_1 \orop \se_2$ & $\sat{M}{\tau}{\se_1} \OR \sat{M}{\tau}{\se_2}$\\
	\ruleandM &$\se_1 \andop \se_2$ & $\sat{M}{\tau}{\se_1} \AND \sat{M}{\tau}{\se_2}$\\
	\rulenotM &$\notop \se_1$ & $\satnot{M}{\tau}{\se_1}$\\
	\specialrule{0pt}{0.5pt}{\belowrulesep}
	\ruleshapeM &$\optional{\closed}\optional{\extra P}\teparen{\te}$ 
	& $\sat{\Msat}{\tau}{\te} \AND \props(\Munsat) \subseteq P' \AND $ \\
	&& $\closed \text{ is present} \implies \props(M) \subseteq \props(\te) \cup P'
	$\\
	&& \text{with}\\
	&&$\quad P' \AFF P\ \text{ if } \extra \text{ is present},\  P' \AFF \emptyset\ \text{ otherwise }$\\
	&& $\quad \Msat \AFF \left\{(n, p, o) \in M \mid \exists \tc{p}{z} \in \tcs(\te).\; \satjump{o}{\tau}{\sref z}\right\}$\\
	&&$\quad \Munsat \AFF \left\{(n, p, o) \in M \mid p \in \props(\te)\right\} \setminus \Msat$ \\
	\specialrule{0pt}{0.7pt}{\belowrulesep}
\ruleextseM & $\extends X\; \shape$ &
$\exists M'\; \text{ and }\;\ \exists M_x \text{ for every $x \in \anc(X)$} \text{ such that}$\\
&&$\quad M = M' \uplus \biguplus_{x \in \anc(X)} M_x \AND$\\
&&$\quad \satjump{M'}{\tau}{\shape} \AND$ \\
&&$\quad \text{for every } x \in \anc(X)\DOT$\\ &&$\quad\quad \satjump{M_x}{\tau}{\mainte(x)} \AND$ \\
&&$\quad\quad \restr(x) \text{ present} \implies \satjump{\left(\bigcup_{z \in \anc(x)} M_{z}\right)\!\!}{\tau}{\restr(x)}$ \\
	\bottomrule
\end{tabular}

\end{table}
A node constraint is satisfied by a set $M$ of neighbourhood triples if their common subject satisfies the constraint.
The definitions of boolean operators are as expected.
For a shape (line~\ruleshapeM), we use the following notations: $\props(M) = \{p \mid (n, p, o) \in M\}$ is the set of properties of the triples in $M$, $\tcs(\te)$ is the set of triple constraints sub-expressions of the triple expression $\te$, and $\props(\te) = \{p \mid \tc{p}{\se} \in \tcs(\te)\}$ is the set of properties in the triple constraints of $\te$.
We identify two significant subsets of $M$:
\begin{itemize}
\item 
$\Msat$ are the triples from $M$ that satisfy some triple constraint in $\te$ under $\tau$,
\item 
$\Munsat$ are the triples from $M$ whose property appears in $\te$ but satisfy none of the triple constraints in $\te$.
\end{itemize}
We require that $\Msat$ satisfies the triple expression $\te$, while the triples in $\Munsat$ may use only extra properties from $P$. In particular, $\Munsat$ must be empty if $\extra P$ is not present.
Additionally, if the shape is $\closed$, then $M$ contains only triples whose properties appear in $\te$ or are in the set $P$ of extra properties.

\begin{example}
	Let $\tau$ and $M_\twofour, M_\twofoursix, M_\twoa, M_\twofoura$ be as in Example~\ref{ex:sat-triple-expr} and define the triple expression $\te = \tc{p}{\exeven} \eachop \tc{p}{\exinf}$.
	Then, for the sets used in line~\ruleshapeM{} we have e.g.: $\Msat_\twofoura = \{(n_2, p, 2), (n_2, p, 4)\}$, $\Munsat_\twofoura = \{(n_2, p, \text{"a"})\}$, $\Msat_\twofoursix = M_\twofoursix$ and $\Munsat_\twofoursix = \emptyset$.
	For example, the following hold:
	\begin{align*}
	&\sat{M_\twofour}{\tau}{\teparen{\te}}
		&& \sat{M_\twofoura}{\tau}{\extra\, \{p\}\, \teparen{\te}}
		\qquad\qquad \satnot{M_\twofoursix}{\tau}{\extra\, \{p\}\, \teparen{\te}} \\
	&\satnot{M_\twoa}{\tau}{\teparen{\te}}
		&& \sat{M_\twofoura \cup \{\nQtwo\}}{\tau}{\extra\, \{p\}\, \teparen{\te}}\\
	&\satnot{M_\twofoursix}{\tau}{\teparen{\te}}		
		&& \satnot{M_\twofoura \cup \{\nQtwo\}}{\tau}{\closed\, \extra\, \{p\}\, \teparen{\te}}
	\end{align*}
\end{example}

For the semantics of shapes with extends (line~\ruleextseM), we use the following notations.
If $X \in \LabX$, then $\anc(X) = \bigcup_{x \in X} \anc(x)$ is the union of the ancestor sets of the labels in $X$; remark that $X \subseteq \anc(X)$. 
For an extendable label $x \in \LabX$ and assuming $\sdef(x) = \extends X\ \shape\ \optional{\andop\ \se}$, we denote by $\mainte(x)$ the triple expression of $\shape$, and by $\restr(x)$ the shape expression $\se$ whenever it exists.

For a shape with extends $\extse = \extends X\; \shape$, different parts of $M$ will be used to satisfy $\shape$ and the shape expressions extended in $\extse$.
First, $M$ must be partitioned as $M' \uplus M_{x_1} \uplus \cdots \uplus M_{x_k}$ where the $x_i$ are the strict ancestors of $\extse$, i.e. $\{x_1, \ldots, x_k\} = \anc(X)$ (for some $k \in \bbbn$).  
Assume also that $\sdef(x_i) = \extends X_i\; \shape_i\; \optional{\andop\; \se_i}$ for every $i \in 1..k$. 
Then we require that:
\begin{itemize}
	\item\label{item:matchvssat} $M'$ satisfies $\shape$ and for every $i \in 1..k$, $M_{x_i}$ satisfies the triple expression $\mainte(x_i)$ which is the triple expression of the shape $\shape_i$;
	\item if present, the restriction $\se_i$ must be satisfied not by $M_{x_i}$ alone, but by the union of all the $M_{z}$ such that $z$ is ancestor of $x_i$ (recall that $x_i \in \anc(x_i)$).
\end{itemize}

\begin{example}
Let $\LabY = \{\exeven, \exinf, \exsup\}$ and let $\tau$, $M_\twofour, M_\twoa, M_\twofoursix, M_\twofoura$ be as in Example~\ref{ex:sat-triple-expr}. 
Consider the schema $\sch = (\LabY, \{x_1, \ldots, x_6\}, \sdef, \emptyset)$, where $\sdef$ is given by:
$$
\begin{array}{ll}
	x_0 \to \extends \_ \teparen{\tc{p}{\exeven}} 
		& \\
	x_1 \to \extends\, x_0\, \teparen{\tc{p}{\exeven}}
		& x_2 \to \extends\, x_1\, \teparen{\tc{p}{\exinf}} \\
	x_3 \to \extends\, x_0\, \teparen{\tc{p}{\exeven}} \andop \teparen{\tc{p}{\exsup}{}^\rep}\qquad
		& x_4 \to \extends\, x_6\, \teparen{\tc{p}{\exinf}} \\
	x_5 \to \extends\, x_0\, \extra\, \{p\}\, \teparen{\varepsilon}
		& x_6 \to \extends\, x_3\, \teparen{\tc{p}{\exeven}}
\end{array}
$$
That is, $\sdef(x_1)$ requires exactly two $p$-triples with even values, and $\sdef(x_2)$ requires an additional $p$-triple 
which value 
is less than five.
Then, e.g., $\sat{M_\twofour}{\tau}{\sdef(x_1)}$ and $\sat{M_\twofoursix}{\tau}{\sdef(x_2)}$, but $\satnot{M_\twofour}{\tau}{\sdef(x_2)}$ and $\satnot{M_\twofoura}{\tau}{\sdef(x_1)}$.
In comparison, $\sdef(x_3)$ requires exactly two $p$-triples with even values greater than five, the latter being expressed by the restriction after the $\andop$.
Therefore, none of the sets from Example~\ref{ex:sat-triple-expr} satisfies neither $\sdef(x_3)$, nor $\sdef(x_4)$, under $\tau$.
Finally, $\sat{M_\twoa}{\tau}{\sdef(x_5)}$ because $\sdef(x_5)$ allows the $\extra$ triple $\nTwoA$.
However, $\satnot{M_\twofoura}{\tau}{\sdef(x_6)}$, because $\extra \{p\}$ in the extended shape expression $\sdef(x_5)$ is not considered for satisfying $\sdef(x_6)$, thus nothing allows the triple $\nFourA$.
\end{example}

We will be interested in correct typings only. 
They are coherent w.r.t. the satisfiability relation, and they ensure that an abstract label can be satisfied only through one of its non-abstract descendants.
\begin{definition}[correct typing]
\label{def:correct-typing}
Let $\graph$ be a graph and $\sch$ be a schema. 
A typing $\tau$ of $\graph$ by $\sch$ is \emph{correct} if for every $(n, z) \in \tau$:
\begin{itemize}
    \item if $z \not\in \LabA$, then $\Sat{n}{\tau}{\sdef(z)}$,
    \item if $z \in \LabA$, then there exists $x \in \desc(z) \setminus \LabA$ such that $\Sat{n}{\tau}{\sdef(x)}$.
\end{itemize}
We denote $\Correct(\graph, \sch)$ the set of correct typings of $\graph$ by $\sch$. 
\end{definition}


\subsection{Well-defined schemas}
\label{sec:semantics-well-defined}

Shape expressions can refer to each other or extend one another and this could result in circular definitions.
Well-defined schemas forbid some circular definitions.
We start by defining a dependency graph that captures how shape expressions depend on one another, then use it to define well-defined schemas.

For every shape expression or triple expression $\se$, let $\subexpr(\se)$ be the set of its sub-expressions, viewed as syntactic objects.
Additionally, $\negsubexpr(\se)$ is the set of sub-expressions of $\se$ that appear in $\se$ under an odd number of occurrences of the negation operator $\notop$. 
Let $\sch = (\LabY, \LabX, \sdef, \LabA)$ be a schema.
We define the following four binary relations for labels $z, z' \in \LabX \cup \LabY$:
\begin{itemize}
\item $\depextends_\sch(z, z')$ iff $H_\sch$ has an edge from $z$ to $z'$ or from $z'$ to $z$;
\item $\depsimple_\sch(z,z')$ iff $z'$ appears as a reference in $\sdef(z)$, i.e. $\sref z' \in \subexpr(\sdef(z))$;
\item $\depshapeneg_\sch(z, z')$ \,iff\, $\optional{\closed} \optional{\extra P} \teparen{\te} \in \negsubexpr(\sdef(z))$ and $\sref z' \in \subexpr(\te)$;
\item $\depextraneg_\sch(z, z')$ \,iff\, $\optional{\closed} \optional{\extra P}  \teparen{\te} \in \subexpr(\sdef(z))$, and $\tc{p}{z'} \in \tcs(\te)$, and $p \in P$.
\end{itemize}
The dependencies $\depshapeneg$ and $\depextraneg$ are called \emph{negative} dependencies.
Obviously, $\depextends$ is new for ShEx with inheritance.
Observe that $\depsimple$ subsumes $\depshapeneg$ and $\depextraneg$.
We say that $z$ depends on $z'$ in schema $\sch$ if $\depsimple_\sch(z,z')$ or $\depextends_\sch(z,z')$.
The schema subscript is omitted whenever the schema is clear from the context.

\newcommand{\EXAMPLEDEPENDENCIES}{%
{\setlength{\tabcolsep}{4pt}
	\begin{tabular}{p{0.32\linewidth}|p{0.3\linewidth}|p{0.31\linewidth}}
		$y_1 \to \teparen{\tc{p}{y_2}} \andop \teparen{\tc{q}{y_3}}$&
		$y_4 \to \teparen{\tc{p}{y_5}}  \orop \teparen{\tc{q}{y_6}}$&
		$x_1 \to \extends\, \_\; \teparen{\tc{p}{y_7}}$ 
		\\
		$y_2 \to \teparen{\tc{q}{y_1}}$& 
		$y_5 \to \notop \teparen{\tc{q}{y_4}}$&
		$x_2 \to \extends\, x_1\;\teparen{\tc{p}{y_8}}$
		\\
		$y_3 \to \extra\{r\}\; \teparen{\tc{p}{y_1}}$&
		$y_6 \to \extra\{p\}\; \teparen{\tc{p}{y_4}}$&
		$y_7 \to \notop\; \teparen{\tc{q}{x_2}}$ \\
		&&
		$y_8 \to \cond$ \\
		\midrule
		$\depsimple_{\sch_1}(y_1, y_2)$ $\depsimple_{\sch_1}(y_1, y_3)$ $\depsimple_{\sch_1}(y_2, y_1)$\;\; $\depsimple_{\sch_1}(y_3, y_1)$ 
		&
		$\depsimple_{\sch_2}(y_4, y_5)$ $\depsimple_{\sch_2}(y_4, y_6)$ $\depshapeneg_{\sch_2}(y_5, y_4)$ $\depextraneg_{\sch_2}(y_6, y_4)$
		&
		$\dep_{\sch_3}(x_1, y_7)$ $\dep_{\sch_3}(x_2, y_8)$  $\depshapeneg_{\sch_3}(y_7, x_2)$ $\depextends_{\sch_3}(x_1, x_2)$ $\depextends_{\sch_3}(x_2, x_1)$
		\\
	\end{tabular}}}

\begin{example}
\label{ex:dep-relations}
Let $\sch_1$, $\sch_2$ and $\sch_3$ be the schemas below, from left to right, with respective sets of simple labels $\LabY_1 = \{y_1, y_2, y_3\}$, $\LabY_2 = \{y_4, y_5, y_6\}$ and $\LabY_3 = \{y_7, y_8\}$ and with extendable labels $\LabX_3 = \{x_1, x_2\}$ for $\sch_3$.
Below each schema are enumerated the facts of its dependency relations. 

\medskip
{\small
\noindent\EXAMPLEDEPENDENCIES}
\end{example}

The dependency graph of a schema $\sch$ regroups all these dependency relations and, together with the extension hierarchy, is used to define well-defined schemas.

\begin{definition}[dependency graph]
	Let $\sch = (\LabY, \LabX, \sdef, \LabA)$ be a schema.
	The \emph{dependency graph} of $\sch$ is a directed labelled graph denoted $D_\sch$ whose set of vertices is $\LabY \cup \LabX$ and that has an edge from $z$ to $z'$ labelled $d$ if $d(z,z')$ holds, where $d$ is one of the dependency relations $\depextends_\sch$, $\depsimple_\sch$, $\depshapeneg_\sch$, $\depextraneg_\sch$.
\end{definition}

\begin{definition}[well-defined schema, stratified negation]
\label{def:well-defined-sch}
A schema $\sch$ is \emph{well-defined} if it satisfies these conditions:
\begin{itemize}
	\item $H_\sch$, the extension hierarchy graph of a $\sch$, is acyclic;
	\item in $D_\sch$, no cycle contains a negative dependency edge labelled $\depshapeneg$ or $\depextraneg$. We say in this case that $\sch$ is \emph{with stratified negation}.
\end{itemize}
\end{definition}

\begin{example}
\label{ex:well-defined}
Considering the schemas from Example~\ref{ex:dep-relations}:
$\sch_1$ is well-defined, as the cycles along $y_1, y_2$ and along $y_1, y_3$ are not forbidden;
$\sch_2$ is not well-defined as its dependency graph contains two cycles -- one along $y_4, y_5$, and another one along $y_4, y_6$ -- each containing a negative dependency;
$\sch_3$ is not well-defined as it has a forbidden cycle  $\depshapeneg_{\sch_3}(y_7, x_2)$, $\depextends_{\sch_3}(x_2, x_1)$, $\dep_{\sch_3}(x_1, y_7)$. Note that the cycle won't be present without the $\depextends$ relation; this illustrates that extension and negation cannot be arbitrarily interleaved.
\end{example}

From here we will discuss only well-defined schemas.
For well-defined schemas, satisfiability relations can be effectively computed.
\begin{lemma}
\label{lem:well-defined-terminates}
The evaluation of $\Sat{n}{\tau}{\sdef(z)}$ always terminates, for every schema $\sch = (\LabY, \LabX, \sdef, \LabA)$, graph $\graph$, node $n$ of $\graph$, typing $\tau$ of $\graph$ by $\sch$, and label $z \in \LabY \cup \LabX$.
\end{lemma}

\subsection{Semantics of ShEx Schemas}
\label{sec:maximal-typing}

The remaining of the section is devoted to the construction of the maximal correct typing. 
It is defined w.r.t. a stratification of the schema, which defines stratums (layers) of labels such that circular dependencies can happen only between labels on the same stratum, while negative dependencies can only go from upper to lower stratums.

\begin{definition}[stratification]
Let $\sch  = (\LabY, \sdef, \LabX, \LabA)$ be a well-defined schema.
A \emph{stratification} of $\sch$ is a function $\stratum: \LabY \cup \LabX \to \bbbn$ such that:
\begin{itemize}
\item if there exists a non-empty path from $z$ to $z'$ in $D_\sch$, then $\stratum(z) \ge \stratum(z')$;
\item if there exists in $D_\sch$ a non-empty path from $z$ to $z'$ that contains an edge corresponding to a negative dependency, then $\stratum(z) > \stratum(z')$.
\end{itemize}
\end{definition}

It is well known that stratified negation, i.e. no cycles containing negative dependencies, guarantees the existence of a stratification.
Note also that w.l.o.g. we can consider that the active range of a stratification is an interval of the form $[1..k]$ for some $k \ge 1$.
In order to define $\taumax$, we introduce a series of typings $\tau_1, \ldots, \tau_k$ such that for every $i \in 1..k$, the typing $\tau_i$ contains only shape labels on stratums $j$ such that $j \le i$.
Each of the intermediate typings is maximal w.r.t. set inclusion, in a sense to be made precise shortly. 
Then $\taumax = \tau_k$. 
We start by introducing some notations.
Given a schema $\sch  = (\LabY, \sdef, \LabX, \LabA)$ and a stratification $\stratum$ of $\sch$ with active domain $[1..k]$, we denote $\LabYstratum_{i}$, resp. $\LabXstratum_{i}$, resp. $\LabAstratum_{i}$ the subsets of labels from $\LabY$, resp. $\LabX$, resp. $\LabA$ whose stratum is at most $i$. Formally, $\LabYstratum_{i} = \{y \in \LabY \mid \stratum(y) \le i\}$, $\LabXstratum_{i} = \{x \in \LabX \mid \stratum(x) \le i\}$ and $\LabAstratum_{i} = \LabA \cap \LabXstratum_{i}$.
Additionally, we let $\schstratum_{i}  = (\LabYstratum_{i}, \LabXstratum_{i}, \sdefstratum_{i}, \LabAstratum_{i})$ be the schema $\sch$ restricted to the labels on stratum at most $i$.
For a typing $\tau$ and a set of labels $Z$ we denote by $\trestr{\tau}{Z}$ the restriction of $\tau$ on $Z$, that is, $\trestr{\tau}{Z} = \{(n, z) \in \tau \mid z \in Z\}$.

\begin{definition}[maximal typing]
	\label{def:str-max-typing}
	Let $\sch = (\LabY, \LabX, \sdef, \LabA)$ be a schema,  $\stratum: \LabY \cup \LabX \to [1..k]$ be a stratification of $\sch$ for some $k \ge 1$, and $\graph$ be a graph.
	For every $1 \le i \le k$, let $\tau_i$ be the typing of $\graph$ by $\schstratum_{i}$ defined inductively by:
	\begin{itemize}
		\item $\tau_1$ is the union of all correct typings of $\graph$ by $\schstratum_{1}$,
		\item for every $1 \le i < k$, $\tau_{i+1}$ is the union of all correct typings $\tau'$ which restriction on $\LabYstratum_{i} \cup \LabXstratum_{i}$ is $\tau_i$. Formally, 
		$\tau_{i+1} = \bigcup \left\{\tau' \subseteq \Correct(\graph, \schstratum_{i+1}) \mid \tau'_{\LabYstratum_{i} \cup \LabXstratum_{i}} = \tau_i\right\}$.
	\end{itemize}
	The typing $\tau_k$ is called the \emph{maximal typing} of $\graph$ by $\sch$ with $\stratum$ and is denoted $\taumax(\sch, \stratum, \graph)$.
\end{definition}

\begin{proposition} 
	\label{prop:maximal-typing-is-correct}
	For every schema $\sch$, stratification $\stratum$ of $\sch$, and graph $\graph$, the typing $\taumax(\sch, \stratum, \graph)$ is a correct typing.
\end{proposition}

The maximal typing can be defined independently on a particular stratification, thanks to the lemma below.
Therefore, we denote by $\taumax(\sch, \graph)$ the unique maximal typing of $\graph$ by $\sch$.
\begin{lemma}
	\label{lem:max-typing-independent-stratification}
	For every schema $\sch$, every graph $\graph$ and all two stratifications $\stratum_1$ and $\stratum_2$ of $\sch$, it holds that $\taumax(\sch, \stratum_1, \graph) = \taumax(\sch, \stratum_2, \graph)$.
\end{lemma}


\section{Validation algorithm}
\label{sec:algorithm}

The validation problem for ShEx is, given a graph $\graph$, a schema $\sch$ and a typing $\tau$ of $\graph$ by $\sch$, determine whether $\tau \subseteq \taumax(\sch, \graph)$. 
Intuitively, the latter means that for every $(n, z) \in \tau$, the node $n$ conforms to the shape expression named $z$.
A validation algorithm for ShEx with inheritance can be based on computing the maximal typing, using a standard \emph{refinement} algorithm such as that presented in \cite{Boneva17}.
It goes as follows.
Compute a stratification $\stratum$ for $\sch$, then construct the series of typings $\tau_{i}$ from Def.~\ref{def:str-max-typing}.
Each of the $\tau_{i}$ is obtained by a refinement that starts with $\tau^{\text{c}}_{i}$, a \emph{complete typing for stratum $i$}, and successively removes from it the pairs that are not conformant.
More precisely, $\tau^{\text{c}}_{i} = \tau_{i-1} \cup \{(n, z) \mid \nodes(\graph) \times (\LabY \cup \LabX) \mid \stratum(z) = i\}$.
Then, if $\tau$ is the current version of the progressively refined typing, we repeatedly remove from it pairs $(n, z) \in \tau$ such that $\Satnot{n}{\tau}{\sdef(z)}$, until the typing becomes correct.
This algorithm also works for ShEx with inheritance.
As shown in Lemma~\ref{lem:well-defined-terminates},
$\Sat{n}{\tau}{\sdef(z)}$ can be effectively computed, so validation is decidable.
In \cite{Boneva17}, the authors also present a recursive algorithm that computes only a portion of $\taumax(\graph, \sch)$ that is relevant for the validation task.
This algorithm also works for ShEx with inheritance.
Because the validation algorithms for ShEx with inheritance are essentially the same as for ShEx 2.1, the effort required to adapt the existing ShEx validators is limited.

The inheritance mechanism does not increase the complexity of validation.
As shown in \cite{Staworko15}, validation of ShEx limited to triple expressions only (i.e. the $\sat{M}{\tau}{\te}$ relation) is NP-complete, while the refinement algorithm requires a polynomial number of steps.
It is easy to see that computing the truth value of $\Sat{n}{\tau}{\sdef(z)}$ is in NP. Therefore

\begin{proposition}
Validation for ShEx with inheritance is NP-complete.
\end{proposition}


\section{Discussion}
\label{sec:discussion}

\subsubsection{Comparison with the ShEx specification}%
\label{sec:comparison-with-shex21}
We use two simplifications.
First, in the specification, references and shape expressions are interchangeable, e.g., it is possible to write $\sref y_1 \andop \sref y_2$ or $\teparen{\tc{p}{(\notop \cond_1 \orop \sref x)}}$. 
Following \cite{Boneva17}, we allow references only in triple constraints.
This does not modify the expressive power of the language, yet it simplifies its presentation and formalization.
Second, in the specification, the sets of labels $\LabY$, $\LabX$ and $\LabA$ are not explicitly given. 
The set of extendable labels $\LabX$ is determined by the use of the $\extends$ keyword, and similarly the set of abstract labels $\LabA$ is determined by the use of $\abst$.

\subsubsection{Inheritance-like features in ShEx 2.1}
ShEx 2.1 allows for the expression of inheritance in some cases.
If a shape expression needs to be extended with new properties only, then conjunction alone can be used, as in this example:
$$
\exFigure \to \teparen{\tc{\excoord}{\exCoord}} \qquad\qquad
\exCircle \to \sref \exFigure\; \andop \teparen{\tc{\exradiusprop}{\exfloat}}
$$
A \exCircle{} is thus a \exFigure{} that has coordinates and a radius.
This works because $\teparen{\tc{\excoord}{\exCoord}}$ and $\teparen{\tc{\exradiusprop}{\exfloat}}$ use disjoint sets of properties and do not interact with each other.
In a slightly different example, we have products with numeric reviews and want to extend them with text reviews:
$$
\exProduct \to \teparen{\tc{\exreview}{\exint}\,\rep} \qquad
\exMyProduct \to \sref \exProduct\; \andop \teparen{\tc{\exreview}{\exstr}\,\rep}
$$
\exMyProduct{} cannot be satisfied because a value of a $\exreview$ property is required to be both integer and string.
We actually want to be able to write something like $\exMyProduct \to \sref \exProduct \eachop \teparen{\tc{\exreview}{\exstr}\,\rep}$, i.e. use the `$\eachop$' operator to extend the contents of \exProduct{} with additional properties.
This would be possible in ShEx~2.1 if the definition of \exProduct{} is a shape (non-terminal $\shape$ in Def.~\ref{def:shape-expr}), which is an important limitation.
Furthermore, such a mechanism would not allow for easy restriction of the inherited shape, for instance if \exMyProduct{} required that numeric reviews have values between 1 and 5.
Overall, ShEx~2.1 or mild extensions of it could provide inheritance-like features with the cost of modelling gymnastics. 

\subsubsection{Design choices} We impose a syntactic restriction on extendable shape expressions. 
They must be conjunctions of a shape and a plain shape expression.
That is, we disallow the extension of more general expressions, in particular expressions of the form $\se_1 \orop \se_2$.
In what follows, we explain the rationale behind this design choice.
On the one hand, the motivating use cases clearly demonstrated the need to extend shape expressions of that form, while the need to extend disjunctions was not supported by the use cases.
On the other hand, extending on disjunctions together with multiple inheritance complicates the definition of the satisfaction relation, as illustrated by the following.
In the example from Sect.~\ref{sec:motivating-example}, \exColCircle{} inherits from \exFigure{} in two different ways.
Suppose that the definition was $\exFigure{} \to \teparen{\tc{\excoord}{\exCartesianCoord}} \orop \teparen{\tc{\excoord}{\exPolarCoord}}$.
This raises the question whether we allow a \exColCircle{} to have both Cartesian and polar coordinates, the one inherited from \exColFig{}, the other inherited from \exCircle.
Allowing it would be undesirable.
Disallowing it would complicate the satisfiability relation as it would require (1) recording which one of the disjuncts (here, Cartesian coordinate and polar coordinate) has been satisfied by each subtype (here, by \exColCircle{} and by \exColFig{}), and (2) checking whether all the subtypes satisfy the same disjunct. 
We decided to take a conservative approach and limit the shape expressions that can be extended.

\subsubsection{Limitations}
The inheritance mechanism can be unintuitive in cases in which a node did not satisfy a \SHEX{} before it was extended, but does conform after the \EX{} is extended, as for instance node {\tt f1} and \SHEX{} \exCircle{} from Sect.~\ref{sec:motivating-example}.
The language is backward compatible in the positive case, i.e. all nodes that satisfy a \SHEX{} continue to satisfy it if it gets extended.
Another possible drawback is the need to know all the descendants of a \SHEX{} in order to validate it.
This can limit the modularity of the validation.

\subsubsection{Inheritance in SHACL?}
SHACL is an RDF validation language published as a W3C recommendation \cite{SHACLSpec}, and has been formalized in several scientific works \cite{Corman2018,ParetiKMN20,Bogaert22}, among others.
We are not aware of the development of an inheritance mechanism for SHACL.
We believe that it would require using conjunction, at least for some use cases.
Using the syntax of SHACL from \cite{Corman2018}, the product example can be captured by:
$$
\exProduct \to\;\, \ge_0 \exreview.\exint \qquad
\exMyProduct \to\;\, \exProduct\;\, \land\, \ge_0 \exreview.\exstr
$$
Then every \exMyProduct{} would be a \exProduct.
The following example cannot be handled in the same way:
a \exClient{} \SHEX{} that has an email needs to be extended to \exMyClient{} that has an additional email.
In ShEx with inheritance it is captured by 
$\exClient \to \extends \_ \teparen{\tc{\exemail}{\exstr}}$, 
$\exMyClient \to \extends \exClient\; \teparen{\tc{\exemail}{\exstr}}$.
A SHACL \exMyClient{} defined in a way similar to \exMyProduct{} above would have only one email.

\subsubsection{Related work} \label{RelatedWork}

The semantics of ShEx with inheritance presented here is an extension of that of ShEx~2.1 from \cite{Boneva17}, e.g. based on the existence of a maximal typing.
Overall, the proof follows the same steps, with additional technicality due to the new satisfiability relation $\sat{M}{\tau}{\se}$ on the one hand, and to the fact that checking the satisfiability of a shape with extends requires dereferencing its ancestors (line~\ruleextseM), on the other hand.
The syntax of the language is slightly different from~\cite{Boneva17}.
There, triple constraints use sets of properties instead of a single property, which allows encoding the $\closed$ and $\extra$ modifiers of a shape within its triple expression.
We cannot rely on the same simplification as the $\closed$ and $\extra$ modifiers of the ancestors are ignored, only the triple expression $\mainte(x)$ is used for satisfaction (line~\ruleextse{} in Table~\ref{fig:sem-se-n}).

The notion of inheritance and its different incarnations has a long tradition in object-oriented programming~\cite{Taivalsaari1996}, under various names such as sub-typing, generalization/specialization, etc. Many of these proposals date to the 80s.
The Liskov substitution principle was proposed in 1987 as a mechanism for conceptual inheritance~\cite{Liskov1987} and the problem of name-collision in multiple-inheritance systems was already discussed at that time~\cite{Knudsen88}. 
The notion of inheritance and inclusion polymorphism is a key feature of object-oriented languages~\cite{Cardelli1985}. 
We follow this tradition by allowing a hierarchy of \SHEX{}s which is similar to sub-typing; we avoid the term subtype to prevent confusion with the notion of types identified by \texttt{rdf:type} in the RDF context. 

Given that ShEx was inspired by XML Schema, the basic functionality of \emph{extends} was also inspired by the extension behaviour in XML Schema~\cite{Simeon2003}, 
but we added multiple-inheritance and removed the requirement of type annotations in data.
PG-Schema~\cite{Angles2023} has been proposed as a schema language for labelled property graphs and includes a notion of inheritance with support for multiple-inheritance. 
However, it checks that nodes conform exactly with their types, without considering their descendants as in our proposal.
Extending ShEx to describe property graphs or RDF-Star has been explored in~\cite{Labra24}, although that paper does not include the inheritance mechanism presented here.

\subsubsection{Conclusions}

We presented an extension of ShEx with an inheritance mechanism.
It is called for by use cases and inspired by inheritance in programming languages.
Its semantics is well-founded, and builds upon the semantics of ShEx~2.1, which makes it easy to integrate into existing ShEx validators.
The extension has several implementations \cite{companion-github-repo}.
The design of the inheritance mechanism required some decisions in order to ensure a good trade-off between keeping its conceptual complexity reasonable, while satisfying the motivating use cases. 
While our proposal has some limitations, we believe that it would allow for better reusability of ShEx schemas. 

\subsubsection{\ackname}
  The work from Jose E. Labra has been partially funded by project ANGLIRU: Applying kNowledge Graphs to research data interoperabiLIty and ReUsability, code: PID2020-117912RB from the Spanish Research Agency and by the regional project SEK-25-GRU-GIC-24-089.

\grayed{

}

\appendix
\section*{Appendix}
\label{sec:appendix-sketches}

We view the satisfiability relations from \AlgoFigures{} as boolean functions.

\vspace{-0.4cm}
\subsubsection{Lemma~\ref{lem:well-defined-terminates} (proof sketch)}
The proof goes by induction on the tree of recursive calls resulting from the evaluation of $\sat{n}{\tau}{\sdef(z)}$, let $\theta$ be that tree.
In $\theta$, every vertex corresponds to one of the lines~\rulefirst--\rulelast{} of the satisfaction functions in \AlgoFigures{} depending on the function being called in that vertex, and on the syntactic form of the shape or triple expression passed as parameter.
Remark that almost all direct recursive calls are made on strict sub-expressions, except for lines~\ruleshape, \ruleextse{} and the two recursive calls $\sat{M_x}{\tau}{\mainte(x)}$ and $\sat{\left(\bigcup_{z \in \anc(x)} M_{z}\right)}{\tau}{\restr(x)}$ on line~\ruleextseM.
Line~\ruleshape{} directly calls line~\ruleshapeM{} which in turns makes recursive calls on strict sub-expressions.
Line~\ruleextse{} directly calls line~\ruleextseM{} that we discuss below.
The above two cases from line~\ruleextseM{} are those that require attention.
We show that in $\theta$, if a vertex is a call $r = \sat{n}{\tau}{\extends X \shape \optional{\andop \se_1}}$ for line~\ruleextse{}, then none of its direct recursive calls mentioned earlier has a descendant equal to $r$.
This is due to the acyclic nature of the extension hierarchy. 
As a consequence, the tree $\theta$ of recursive calls is finite.

\bigskip
\noindent
\textbf{Proposition~\ref{prop:maximal-typing-is-correct}} The proof goes by induction on the number of stratums.

\smallskip
\noindent
\textit{Base case} There is a unique stratum and the dependency relations $\depextraneg_{\sch}$ and $\depshapeneg_{\sch}$ are empty.
As a consequence, (*) every sub-expression $\notop s_1$ that appears in the schema is s.t. $s_1$ contains no references, and every sub-expression $\optional{\closed}\optional{\extra P} \teparen{\te}$ is s.t. for every $\tc{p}{\sse} \in \tcs(e)$, if $\sse$ is a reference, then $p \not \in P$. 

We show that (**) if $\tau'$ and $\tau''$ are correct typings, their union $\tau = \tau' \cup \tau''$ is also a correct typing then, because there is a finite number of typings, the union of all correct typings (i.e. $\taumax$) is also correct.
For (**), it is enough to show that if $\sat{n}{\tau'}{\sdef(z)}$\footnote{the case for $\tau''$ is symmetric} evaluates to true, then $\sat{n}{\tau}{\sdef(z)}$ evaluates to true, for every $n$ and every $z \not\in \LabA$
s.t. either $(n, z) \in \tau$, or there exists $x \in \LabA$ with $z \in \desc(x) \setminus \LabA$ and $(n, x) \in \tau$.
This goes by structural induction on the tree of recursive calls for the evaluation of $\sat{n}{\tau'}{\sdef(z)}$, let $\theta$ be that tree.

Once again, we distinguish the cases by the corresponding line in \rulefirst--\rulelast{} of the algorithms.
Lines~\ruleepsilon, \rulecond, \ruleref, \rulecondM{} are base cases (leaves of $\theta$). Lines~\ruleepsilon, \rulecond{} and \rulecondM{} do not rely on the typing, while for line~\ruleref{}, obviously $(n', z') \in \tau' \implies (n', z') \in \tau$ for all $n', z'$.
For the induction case, we consider a vertex $r' = \sat{\nM}{\tau'}{\se}$ of $\theta$ (where $\nM$ is the node or set of triples parameter of the call), and the following induction hypothesis:
(ih) for every $r''$ strict descendant of $r'$ in $\theta$, if $r''$ evaluates to true, then $r''$ with same parameters except for the typing $\tau'$ being replaced by $\tau$ also evaluates to true.
We show that then $\sat{\nM}{\tau}{\se}$ evaluates to true.

Lines \ruletc{}, \ruleshape{}, \ruleextse, \ruleone, \ruleor, \ruleand, \ruleorM, \ruleandM{} easily follow from (ih).
For lines~\ruleeach, \rulerep, \ruleextseM{} we can take the same values for the sets $M_i$, $M_x$ and $M$ and then it again follows from (ih).
For lines~\rulenot{} and~\rulenotM, the expression of the call is $\notop s_1$ and then by (*), $s_1$ does not contain references, so its satisfiability does not depend on the typing.
Therefore, the only challenge is line~\ruleshapeM.
So, consider a call $r = \sat{M}{\tau}{\optional{\closed} \optional{\extends P} \teparen{\te}}$, let $r' = \sat{M}{\tau'}{\optional{\closed} \optional{\extends P} \teparen{\te}}$, and let:
\begin{itemize}
	\item $\Msat_{\tau'} = \left\{(n, p, o) \in M \mid \exists \tc{p}{\se} \in \tcs(\te).\; \sat{o}{\tau'}{\se}\right\}$, and
	\item $\Msat_\tau = \left\{(n, p, o) \in M \mid \exists \tc{p}{\se} \in \tcs(\te).\; \sat{o}{\tau}{\se}\right\}$
\end{itemize} 
be the corresponding sets computed in the calls $r'$ and $r$, respectively.
We show that if $r'$ evaluates to true, then $\Msat_{\tau'} = \Msat_\tau$, which using (ih) directly allows deducing that $r$ evaluates to true.

So, assume that $r$ evaluates to true.
Remark first that by (ih) we have $\Msat_{\tau'} \subseteq \Msat_\tau$.
Suppose by contradiction that $(n_1, p, n_2) \in M$ and $\tc{p}{\se} \in \tcs(e)$ are s.t. $\sat{n_2}{\tau}{\se}$ but $\satnot{n_2}{\tau'}{\se}$; thus $(n_1, p, n_2) \in \Msat_\tau \setminus \Msat_{\tau'}$.
Then by (*), we deduce that $\se$ is of the form $\sref z'$, so $p \not\in P$.
Then $(n_1, p, n_2)$ belongs to the set $\Munsat_{\tau'} = \{(n_1, q, n') \in M \mid q \in \props(\te)\} \setminus \Msat_{\tau'}$ computed during the evaluation of $r$, so $\props(\Munsat_{\tau'}) \not\subseteq P$, which contradicts the fact that $r$ evaluates to true. 

\smallskip\noindent
\textit{Induction case}
Let $i > 1$ be in the range of $\stratum$, the induction hypothesis is (IH) for every $j < i$, $\tau_{j}$ is a correct typing.
We again show that if $\tau', \tau''$ are correct typings for $\schstratum_{i}$, their union $\tau = \tau' \cup \tau''$ is also a correct typing. 
Thus, we need to show that 
if $\sat{n}{\tau'}{\sdef(z)}$ evaluates to true, then $\sat{n}{\tau}{\sdef(z)}$ evaluates to true, for every $n$ and every $z \not\in \LabA$
s.t. either $(n, z) \in \tau$, or there exists $x \in \LabA$ with $z \in \desc(x) \setminus \LabA$ and $(n, x) \in \tau$. 
We assume that $\stratum(z) = i$, otherwise (IH) applies directly.
Similarly to the base case above, the proof goes by structural induction on the tree of recursive calls $\theta$ for $\sat{n}{\tau'}{\sdef(z)}$.
The challenging cases are lines~\rulenot, \rulenotM{} and \ruleshapeM.
We do not have any more the property (*) from above allowing to easily deal with lines~\rulenot{} \rulenotM.
However, we can show that if $\theta$ contains a node with expression $\notop \se_1$, then its evaluation depends only on labels lying on stratums strictly less than $i$. 
Moreover, by definition of stratification, all labels that appear in $\se_1$ are necessarily on some stratum that is strictly less than $i$.
Note however that, because of the extension mechanism and line~\ruleref, the evaluation of $\se_1$ might depend on labels that are descendants of a label appearing in $\se_1$.
Because of line~\ruleextseM, it can also depend on labels that appear in the definitions of ancestors of $z$. 
Thanks to the $\depextends$ dependency relation, such labels are also on stratums strictly less than $i$.
We thus show that $\sat{\nM}{\tau'}{\notop \se_1}$ evaluates the same as $\sat{\nM}{\tau'_{|i-1}}{\notop \se_1}$, where $\tau'_{|i-1}$ is the restriction of $\tau'$ to labels on stratum at most $i-1$.
Similarly, $\sat{\nM}{\tau}{\notop \se_1}$ evaluates the same as $\sat{\nM}{\tau_{|i-1}}{\notop \se_1}$.
By definition we know that $\tau'_{|i-1} = \tau_{|i-1} = \tau_{i-1}$, then $\sat{\nM}{\tau'}{\notop \se_1}$ and $\sat{\nM}{\tau}{\notop \se_1}$ have the same value.

As for line~\ruleextse, the proof uses the arguments from above regarding negative dependencies and stratification, and the arguments from the same line in the base case.

\vspace{-0.35cm}
\subsubsection{Lemma~\ref{lem:max-typing-independent-stratification} (proof sketch)}
Let $\sigma'$ be a most refined stratification obtained by taking every strongly connected component of $H_\sch$ as a separate stratum.
Then for every stratification $\stratum$, every stratum of $\stratum$ is a union of stratums of $\stratum'$.
Using this property, we can show that for every stratification $\stratum$, $\taumax(\sch, \sigma, \graph) = \taumax(\sch, \sigma', \graph)$.
It then immediately follows that $\taumax(\sch, \sigma_1, \graph) = \taumax(\sch, \sigma_2, \graph)$.

\bibliographystyle{splncs04}
\bibliography{main}


\end{document}